\documentclass[aps,prl,showpacs,showkeys,twocolumn,superscriptaddress,
	      floats,floatfix,preprintnumbers]{revtex4}
\usepackage{amssymb,amsmath}
\usepackage{bm,url,graphics,subfigure,epsfig}
\usepackage{ulem}
\usepackage{color}

\def \obyk {\frac{\omega}{k}}
\def \qbym {\frac{q}{m}}
\def \xx {\bm{x}}
\def \vv {\bm{v}}

\def \cE {\mathcal{E}}
\def \Np {N_{\rm p}}

\def \curl {{\bm \nabla} \times}

\newcommand{\deldelt}[1]{\frac{\partial #1}{\partial t}}

{}

\def \omc  {\omega_{\rm c}}
\def \Bzero  {B_{\rm 0}}
\def \rc  {r_{\rm c}}

\def \dx    {\delta x}

\def \dxx   {\delta\bm x}
\def \dvv   {\delta\bm v}

\def \BB {\bm B}
\def \EE{\bm E}

\def \curl {{\bm \nabla}\times}

 \def \bb {{\bm B}}

\newcommand{\nab}{\mbox{\boldmath $\nabla$} {}}
\newcommand{\eq}[1]{(\ref{#1})}

\def \aap {Astron. Astrophys. }

\def \apj {Astrophys. J.}

\begin{document}
\title{Particle energization through time-periodic helical magnetic fields}
\author{Dhrubaditya Mitra}
\affiliation{Nordita, KTH Royal Institute of Technology and Stockholm University,
Roslagstullsbacken 23, 10691 Stockholm, Sweden}
\author{Axel Brandenburg}
\affiliation{Nordita, KTH Royal Institute of Technology and Stockholm University,
Roslagstullsbacken 23, 10691 Stockholm, Sweden}
\affiliation{Department of Astronomy, AlbaNova University Center,
Stockholm University, 10691 Stockholm, Sweden}
\author{Brahmananda Dasgupta}
\affiliation{Center for Space Plasma and Aeronomic Research, University of Alabama in Huntsville, USA}
\author{Eyvind Niklasson}
\affiliation{Nordita, KTH Royal Institute of Technology and Stockholm University,
Roslagstullsbacken 23, 10691 Stockholm, Sweden}
\author{Abhay Ram}
\affiliation{Plasma Science and Fusion Center, Massachusetts Institute of Technology, Cambridge, MA 02139, USA}
\date{\today,~  }
\begin{abstract}
We solve for the motion of charged particles in a helical time-periodic ABC (Arnold-Beltrami-Childress)
magnetic field.
The magnetic field lines of a stationary ABC field with coefficients $A=B=C=1$ 
are chaotic, and we show that the motion of a charged particle in such a field
is also chaotic at late times with positive Lyapunov exponent. 
We further show that in time-periodic ABC fields, the kinetic energy of a charged particle can 
increase indefinitely with time. 
At late times the mean kinetic energy grows as a power law in time with an exponent
that approaches unity.
For an initial distribution of particles, whose kinetic energy is uniformly distributed
within some interval, the PDF of kinetic energy is, at late times,
close to a Gaussian but with steeper tails. 
\end{abstract}
\keywords{Particle acceleration, Cosmic Rays, Dynamical Systems}
\pacs{
94.20.wc, 
96.50.Pw, 
98.70.Sa 
}
\preprint{NORDITA-2013-38}
\maketitle
\section{Introduction}

Production of charged particles with energies far exceeding the thermal energy is known to 
be a very common  phenomenon in cosmic plasma. Such energetic particles, which include
interplanetary, interstellar and galactic cosmic rays,  are believed to be  
produced in various astrophysical bodies, from magnetospheric to cosmic plasmas, including solar flares, 
coronal mass ejections (CME) and supernova remnants.
In other words, acceleration of charged particles occurs ubiquitously in the plasma Universe. 
Investigations of particle energization remains a major topic of astrophysics.
The seminal paper in this field is by Fermi~\cite{fer49}, who proposed that charged particles in cosmic
rays can attain very high energies by being repeatedly reflected by two magnetic mirrors moving 
towards each other.

Fermi's idea of energization was tested in a simple setting in 
the now-famous Fermi-Ulam model \citep{ula61} whose numerical simulations 
showed that, although the motion of a particle
reflected by moving walls can be chaotic, on average no energy is gained by the particle
if the motion of the moving wall is a smooth function of time.
This result was elaborated upon in the early days of 
research in nonlinear dynamical systems~\cite{zas+chi65,bra71,lie+lic72,lic+lie83}, 
to show that energy can grow as a power law in time if the motion of
the wall is not smooth in time, e.g., random or a saw-tooth profile in time. 
The Fermi-Ulam problem in more than one dimension, sometimes called 
``billiard problems with breathing walls'', 
allow energization of particles; see, e.g., \cite{gel+rom+tur12} for a recent review.
In some specially constructed cases, even exponential-in-time energy growth is 
possible~\cite{gel+rom+sha+tur11}. 
Such problems, although of fundamental interest,
are somewhat removed from the problem of energization of
charged particles in time-dependent magnetic fields. 
In this paper, we show that the energy of charged particles can increase as a power law in time
in a simple helical magnetic field, whose components are slowly varying sinusoidal functions 
of both space and time.

The motion of a charged particle in deceptively simple magnetic fields can
be very complex, even chaotic.
A recent paper~\cite{ram+das10} has shown that
very simple current configurations, for example a circular current loop in the $x$-$y$ plane
plus a line current along the $z$ axis passing through an off-center point, can give rise to
a magnetic field whose magnetic lines of force are non-integrable and 
chaotic \footnote{Of course, charged particles in a magnetic field do not move along magnetic
field lines, and their motion may be chaotic even in magnetic fields where the lines of force are
integrable.}.
The motion of a charged particle in such a chaotic magnetic field may or may not be
chaotic. 
Recently, it was conjectured \cite{das+li+li+ram12} that,
if such a chaotic magnetic field changes with time, it may be able
to impart significant energy to a charged particle. 
Our work, described below, demonstrates the feasibility of effective energization of 
charged particles by a time-varying chaotic magnetic field. We show that this process can lead to 
an indefinite energization of a charged particle to relativistic energies, given enough time. 

\section{Model}

It is now generally accepted that astrophysical magnetic fields are generated by some
dynamo mechanism, i.e., a mechanism by which the kinetic energy of
the fluid is converted to magnetic energy~\cite{BS05}. 
If the characteristic length scale of the magnetic field, generated by
the dynamo mechanism, is larger than the energy-containing scales of the fluid, the dynamo is called a
large-scale dynamo. The most common examples of astrophysical magnetic fields, e.g.,
galactic magnetic fields, solar magnetic fields
and planetary magnetic fields are generated by a large-scale dynamo.
Almost all large-scale dynamo mechanisms demand that the fluid flow
possesses helicity, i.e., handedness.
The helicity of the flow is often described by the well-known
$\alpha$ effect which was proposed by Parker \cite{par55} with a
detailed mathematical basis provided by Steenbeck, Krause and R\"adler~\cite{kra+rad+ste71}.
The helical flow typically generates a helical field. Thus almost all
large-scale astrophysical magnetic fields are helical in nature. 
This is also true of large-scale magnetic fields generated by numerical simulations \cite{B01};
see also \cite{BS05} and references therein.
Observations of solar flares also show the helical nature of
magnetic field ejected from the Sun \cite{dem+man+van+tho+plu+kov+aul+you02}.
The helical nature of the magnetic field carried by the solar wind has
also been observed \cite{bra+sub+bal+gol11}.
One of the simple examples of a helical field, which is also a force-free
field, is the Arnold-Beltrami-Childress (ABC) flow.
The streamlines of the ABC flows are chaotic~\cite{dom+fri+hen+gre+sow86}. 

Here we study the possibility of energization of a test-particle in a 
magnetic field, which is time-dependent ABC function:
\begin{equation}
\BB = \bb\sin\omega t,
\end{equation}
where $\bb$ is an ABC function with wavenumber $k$:
\begin{eqnarray}
\bb_x &=& \Bzero(A\sin kz + C\cos ky),\nonumber \\
\bb_y &=& \Bzero(B\sin kx + A\cos kz),\nonumber \\
\bb_z &=& \Bzero(C\sin ky + B\cos kx).
\label{eq:abct}
\end{eqnarray}
Here we choose $A=B=C=1$ (which is the case where the ABC flow has chaotic streamlines) ; where the 
time-dependent part is a sine function with circular frequency $\omega$ and $t$ is time.
Although there could be time variations with multiple time scale are
expected to occur in a realistic astrophysical environment, in this
work we have chosen, as an example,  a simple sinusoidal time variation of the helical field.

Furthermore, the energization  of test-particles has been observed in complex (in space, but constant in time) turbulent 
electric fields generated  by time variation of fluctuating magnetic field in 
direct numerical simulations of  MHD~\cite{ono+isl+vla06,dmi+mat+sen04}.  
It then behooves us to ask the question, what are the essential ingredients of the
energization process?
Can a simple magnetic field like an ABC field that is periodic in both 
space and time, but is nevertheless expected to give rise to chaotic motion of test-particles,
energize test-particles indefinitely?
Finally,  the ABC field is an eigenfunction of the curl operator
hence it is easy to solve Maxwell's equations
to obtain the electric field generated by the time-dependent part of the ABC field.

Through Maxwell's equations, the time-varying ABC field shall generate
a fluctuating electric field, given by 
\begin{eqnarray}
\curl \EE &=& - \deldelt{\BB} \nonumber \\
          &=& -\bb\omega\cos\omega t.
\label{eq:max}
\end{eqnarray}
Here, $\EE$ is the electric field, which is given by
\begin{eqnarray}
\EE = \obyk\bb \cos\omega t.
\label{eq:abcE}
\end{eqnarray}
In addition, we assume that $\omega/k \ll c$, where $c$ is the speed of light.
Hence we can safely ignore the displacement current in Maxwell's equations
of electrodynamics. We further assume the particle to be non-relativistic
hence its  equations of motion are given by Newton's second law of motion
\begin{eqnarray}
\dot{\xx} &=& \vv, \nonumber \\
\dot{\vv} &=& \qbym\left(\EE + \vv \times \BB\right).
\label{eq:newton}
\end{eqnarray}
Here and henceforth, the dot denotes a time derivative. 
In what follows, unless otherwise stated, length scales are normalized by $1/k$, time by
$\omc$ where $\omc$ is defined to be the  characteristic gyro-frequency,
\begin{equation}
\omc=\frac{q \Bzero}{m}.
\end{equation}
We solve for $\Np=409600$ copies of this six dimensional dynamical
system. 
Initially, the particle positions are uniformly distributed inside a cube of 
dimensions $2\pi\times2\pi\times2\pi$ with velocity along the $x$-axis
uniformly distributed between $-0.01$ to $0.01$  and $k=1$.
We use a fourth order Runge-Kutta method with fixed step size \cite{Pre+Fla+Teu+Vet92} as our
time-stepping algorithm.
This algorithm is not energy-conserving by construction. However we have checked that,
in practice, energy conservation is satisfied with a high degree of 
accuracy \footnote{For the case of stationary magnetic field, in which case energy 
should be conserved, the fractional change in energy over the whole period of 
integration is negligible; less than $10^{-9}$.}.   
We have also checked for representative runs that a Runge-Kutta-Fehlberg~\cite{Pre+Fla+Teu+Vet92} scheme with
variable step-size gives the same results. 
The computation are done with a python package
to  solve ordinary differential equations, and the figures are prepared using Matplotlib~\cite{hun07}. 

\section{Results}
\begin{figure}[h]
\begin{center}
\includegraphics[width=0.98\columnwidth]{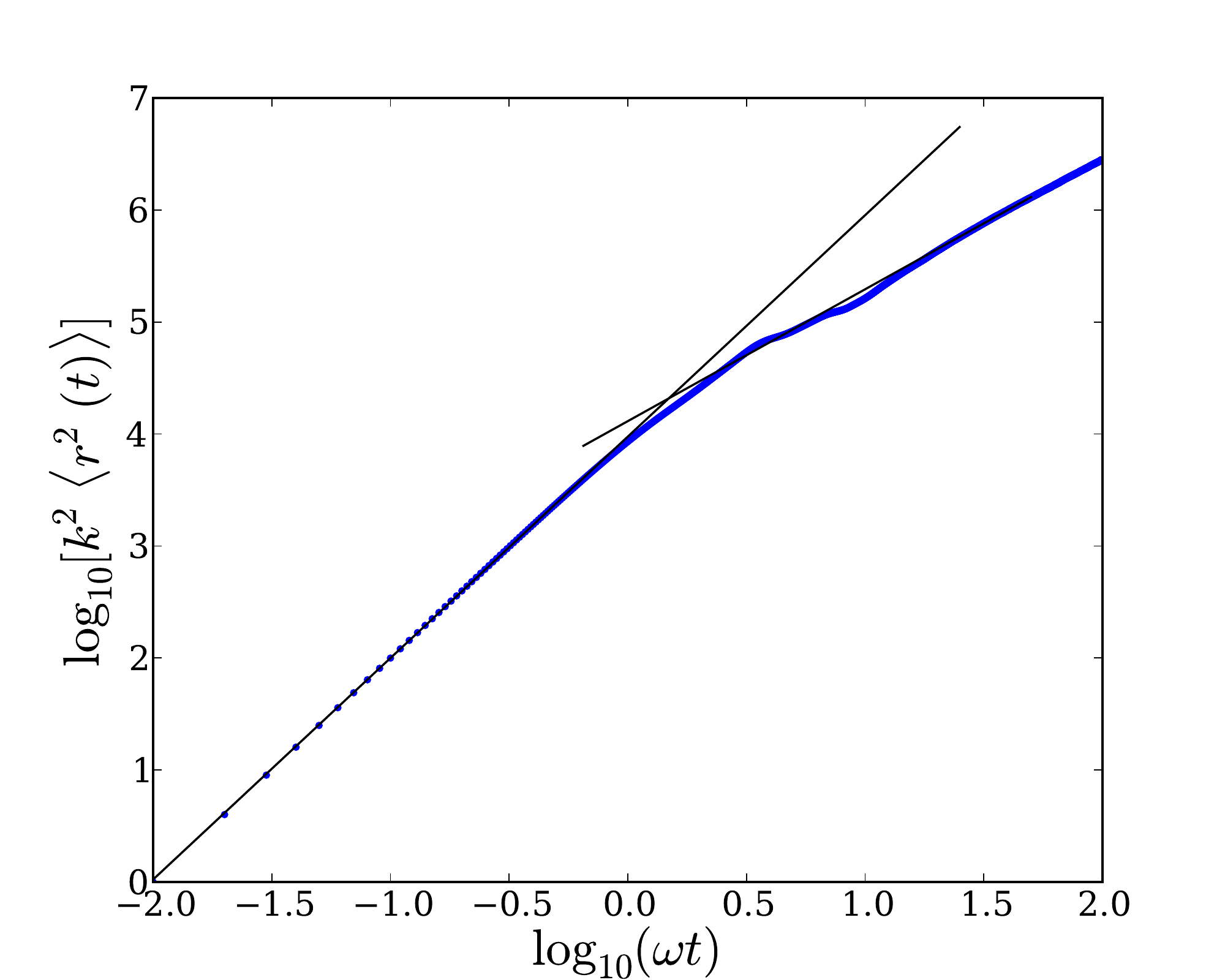}
\caption{\label{fig:r2vst_B0}
Average displacement $\langle r^2 \rangle$ as a function of time
for particles in a stationary ABC magnetic field ($\omega=0$).
The ordinate is normalized with $k^2$ and the abscissa with the characteristic 
gyro-frequency $\omc$. 
The two straight lines to the left and right of the curve are fits
with slopes $1.9$ and $1.1$, respectively.
}\end{center}
\end{figure}

Let us first study the properties of trajectories of particles for the
case in which the magnetic field is constant in time.
In this case  we define the characteristic gyroradius of the particle,
\begin{equation}
\rc = \frac{v}{\omc},
\end{equation}
where $v$ is the magnitude of the velocity of the particle, and $\omc$  the 
characteristic gyro-frequency,
is constant in time, because the energy of the particle is conserved.  Let us first consider the case
$\rc \ll 2\pi/k$ with $k$ being the characteristic wavenumber of the magnetic field. 
In this case, for times $t \ll t_{\rm trans} = 1/(kv)$, with $v=|\vv|$, the particles move 
 in a field that is almost a constant and hence its motion is not random.
Here, $t_{\rm trans}$ is the time it takes a typical particle to go a distance equal to 
the wavelength of the magnetic field.
The net displacement of the particle is due to  the component of its velocity 
parallel to the local magnetic field. This component of velocity
is also constant.  
Hence, the mean-square displacement of the particles obeys
\begin{equation}
\langle r^2 \rangle \sim t^2 \quad \mbox{for $t \ll t_{\rm trans}$}. 
\label{eq:r2vst}
\end{equation}
In the other limit, $\rc \gg 2\pi/k$, the particles encounter
significant change in the magnetic field even within one gyro-orbit.
Hence, we expect non-trivial behavior in this regime.
The mean-square displacement (normalized by $k^2$) as a function of 
time (normalized by $\omc$), in log-log scale, is plotted in Fig.~\ref{fig:r2vst_B0}.
For short times, Eq.~(\ref{eq:r2vst}) is indeed verified;
for large times we find that the mean-square displacement
grows approximately linearly with time (slope of the fit is 1.1),
i.e., Brownian motion is observed.

\begin{figure}[h]
\begin{center}
\includegraphics[width=0.98\columnwidth]{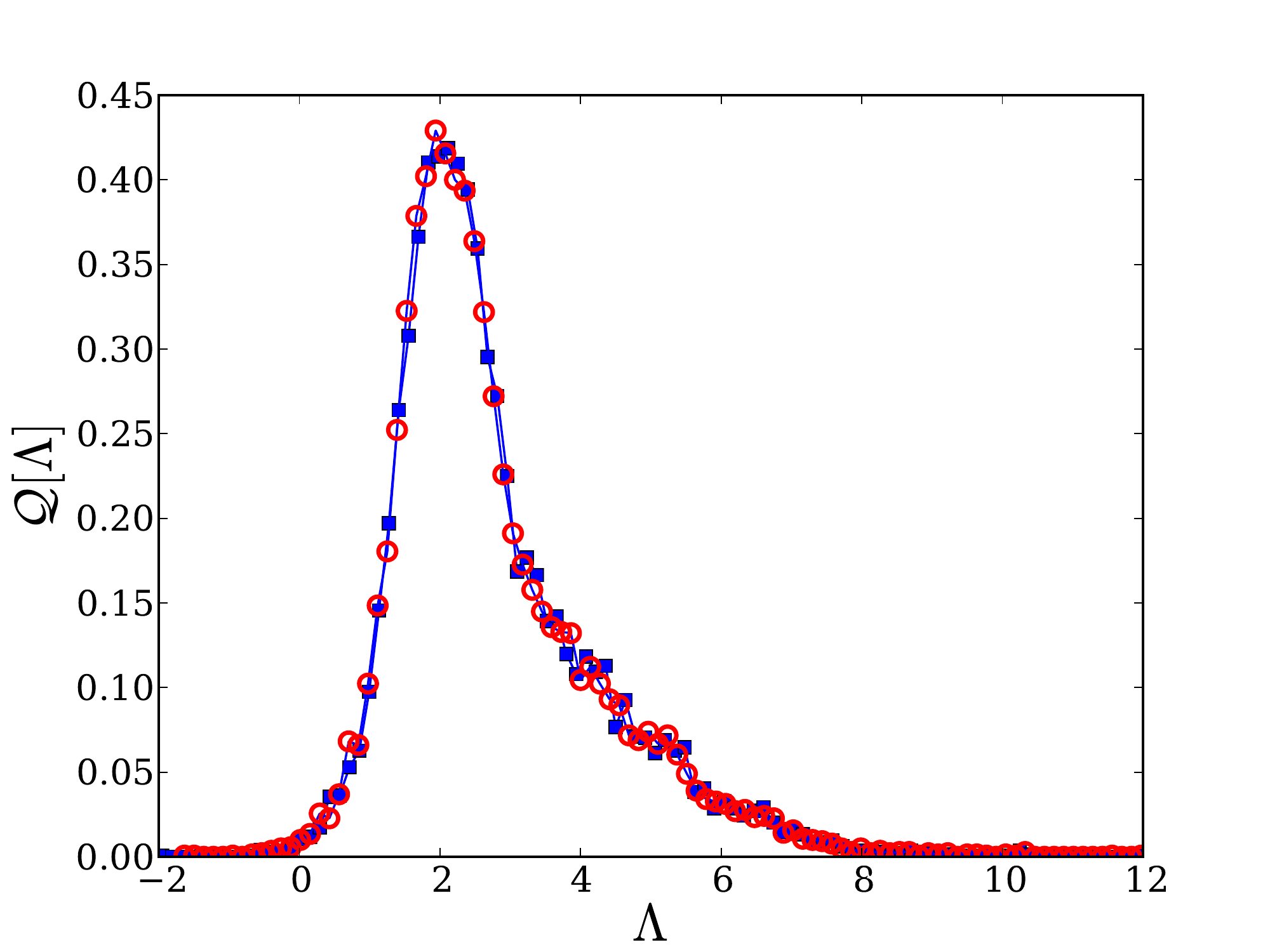}
\caption{\label{fig:FTLE} (Color online)
The PDF $\mathcal{Q}(\Lambda)$ of the
Lyapunov exponents, defined in Eq.~(\ref{eq:lyap}),
for two time differences, $\omega_c \delta t = 130$ (blue, filled squares)
and $140$ (red, open circles). 
}\end{center}
\end{figure}

Whether or not chaotic trajectories exist in this system (i.e., 
with stationary magnetic field)
can be investigated by studying the tangent system,
\begin{eqnarray}
\dot{\dxx} &\!=\!& \dvv, \nonumber \\
\dot{\dvv} &\!=\!& \frac{q}{m} \big[(\dxx\cdot\nab)\EE
+\vv\times(\dxx\cdot\nab)\BB + \dvv\times\BB \big].\quad
\end{eqnarray}
This dynamical system clearly depends on the trajectory in phase space
given by $\xx(t),\vv(t)$.
It is possible to solve this system for each trajectory
given by the solutions of Eq.~(\ref{eq:newton}).
For each such trajectory one can calculate the quantity
\begin{equation}
\Lambda = \lim_{t\to\infty} \frac{1}{t} \ln\frac{|\dx(t)|}{|\dx(0)|},
\label{eq:lyap}
\end{equation}
which is the largest Lyapunov exponent for that particular trajectory.
Each trajectory is then labeled by the initial choice of position and velocity.
For a set of random initial conditions, we have calculated  
the probability density function (PDF), $\mathcal{Q}(\Lambda)$.
In Fig.~\ref{fig:FTLE} we plot  $\mathcal{Q}(\Lambda)$ for two
time differences. 
The two PDFs are quite close, and both of them are Gaussian with 
a positive mean.  
Hence, we conclude that at large times, $\mathcal{Q}(\Lambda)$
is Gaussian with a positive mean, i.e.,
the dynamical system \eq{eq:newton} even with a time-independent magnetic field, which
also implies zero electric field, has chaotic trajectories. 


Now let us study the case with a time-periodic magnetic field. 
We limit ourselves to the case where $\omega/\omc$
is very small and $\omega/k$ remains equal to unity.
In other words, we consider a magnetic field that varies slowly in both space and time. 

\begin{figure}[h]
\begin{center}
\includegraphics[width=0.8\columnwidth]{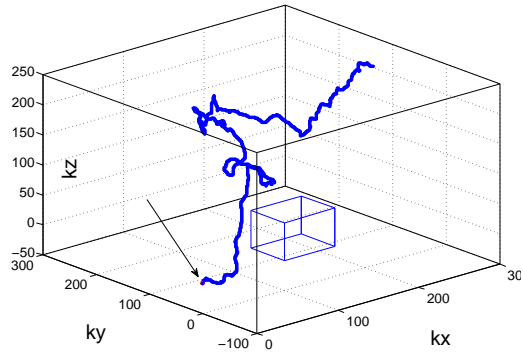}
\caption{\label{fig:track}
(Color online) A typical track of a particle in the magnetic field. The initial position
of the particle is shown by an arrow. The ABC field is periodic over a cube of unit length in 
in units plotted in this figure. To give an idea of scales, a cube with each side equal to $10$ is
shown in the figure.  
}\end{center}
\end{figure}

\begin{figure}[h]
\begin{center}
\includegraphics[width=0.98\columnwidth]{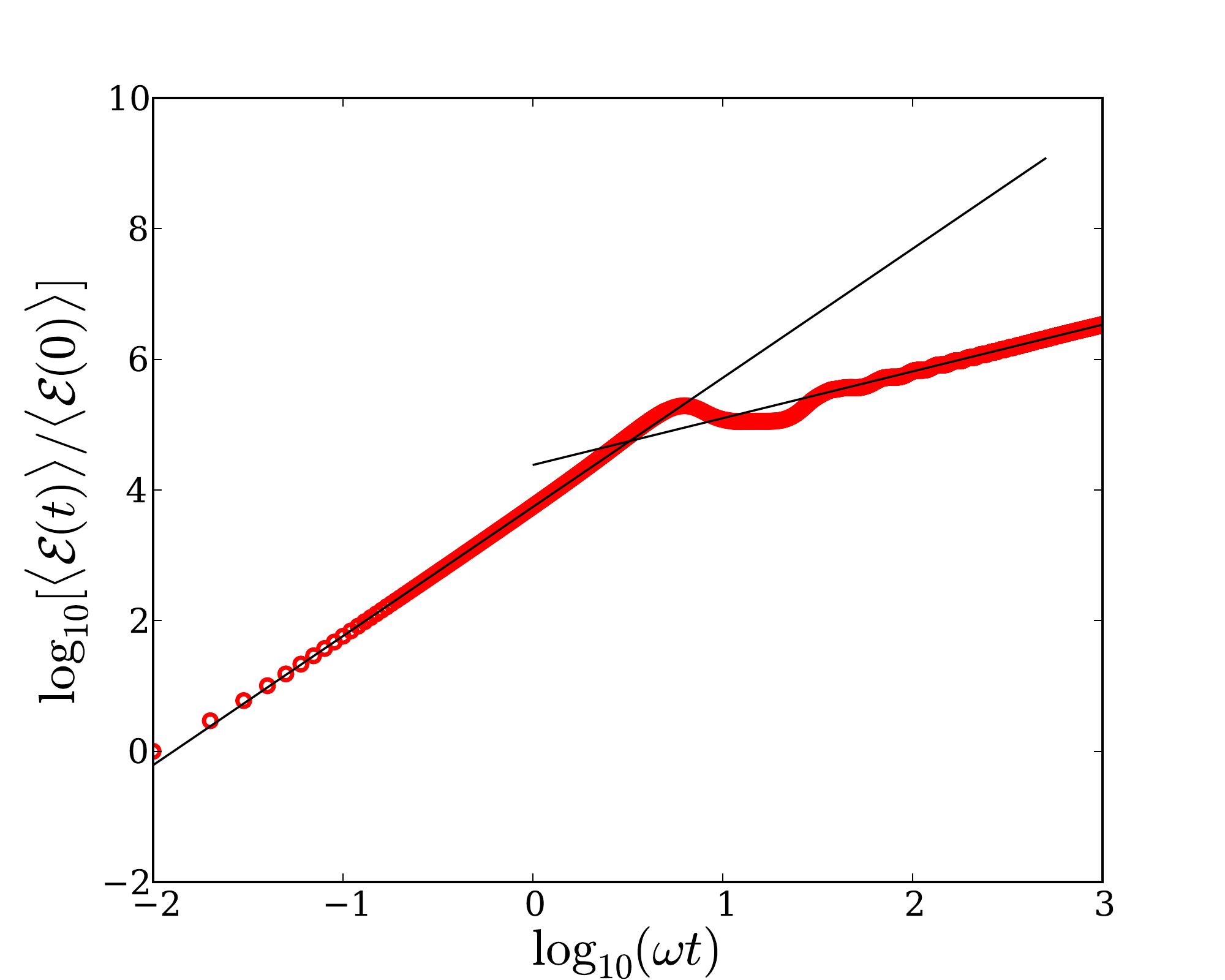}
\caption{\label{fig:evst2}
Mean energy of the particles as a function of time for run ${\tt R2}$; see Table~\ref{table:param}.
The two straight lines are
fits (in least-square sense) with slopes $1.95$ and $0.77$ for small $t$ and large $t$, respectively.
}\end{center}
\end{figure}

\begin{figure}[h]
\begin{center}
\includegraphics[width=0.95\columnwidth]{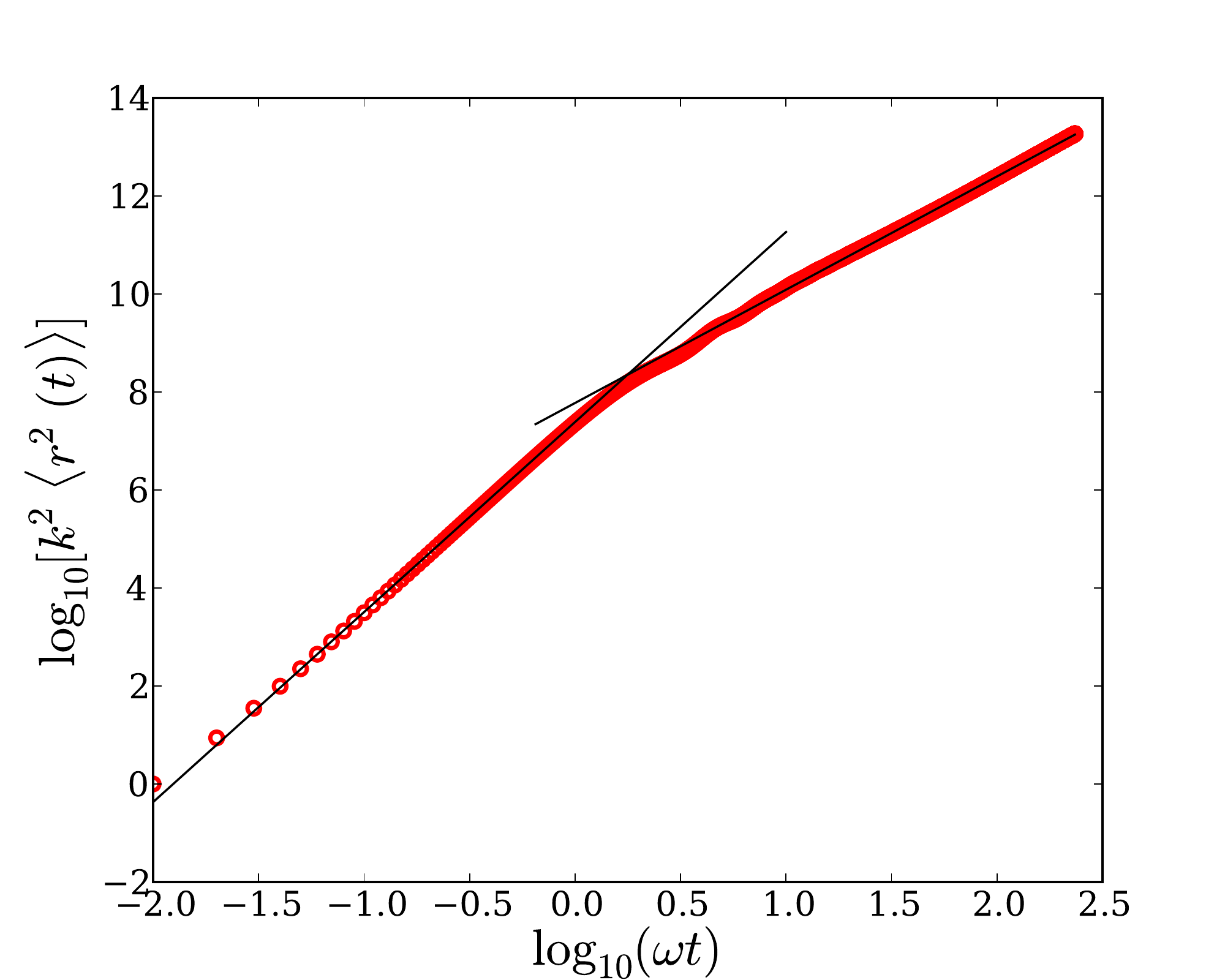}
\caption{\label{fig:r2vst}
Plot of the mean square displacement of the particles as a function of time, for 
run ${\tt R2}$.  The two straight lines are the 
fits (in least-square sense) with slopes $3.9$ and $2.1$
for small $t$ and large $t$, respectively.
}\end{center}
\end{figure}

A typical path is shown in Fig.~\ref{fig:track}. 
Note that, although the magnetic field is periodic with wavenumber $k=1$, the particle
trajectories themselves are not periodic. To illustrate this in Fig.~{fig:track} we 
have also plotted a cube whose each side is $10$ times the length scale over which
the magnetic field is periodic.
In Fig.~\ref{fig:evst2} we show the growth of kinetic energy  (per unit mass) averaged over the
total number of particles,
\begin{equation}
\cE(t) \equiv  \frac{1}{2}  \left \langle v^2 \right \rangle.
\label{eq:energy}
\end{equation}
At short times, $\cE(t)$ behaves as $t^2$, but at later time 
goes like $t^{\xi}$ with an exponent
$\xi$ that is not universal but depends on $\omega$; 
see Figure~\ref{fig:evst2} and Table~\ref{table:param}.
The gyro-radius of the charged particle grows with time as energy grows.
The first stage of the growth, over which
energy grows as $t^2$, continues till the gyroradius becomes of the same order
as the characteristic scale of the ABC field.
Figure~\ref{fig:evst2} shows that the first stage of the growth, over which
energy grows as $t^2$, continues till the gyroradius becomes of the same order
as the characteristic scale of the ABC field.
The growth at these short times can be understood by reminding
ourselves that for $k\rc \ll 1$, a particle does not encounter
significant spatial change in the magnetic field.
As $\omega/\omc$ is small, the particle effectively moves under a
constant force (the electric field), hence its 
energy grows quadratically with time. 
At later times, the growth slows down to $t^{\xi}$ 
where the exponent $\xi$ for different values of $\omega$ is given in table~\ref{table:param}.
We find that, as $\omega/\omc \to 0$, $\xi$ approaches unity. 
This shows that the process we observe can be interpreted as a  sub-diffusive process in
momentum space. 
In real space, we simultaneously find that the mean-square displacement is proportional to
$t^4$ for small time and goes as $t^2$ for large times.
A representative plot of the mean-square displacement as a function of
time is given in Fig.~\ref{fig:r2vst}. 

\begin{figure}[h]
\begin{center}
\includegraphics[width=0.98\columnwidth]{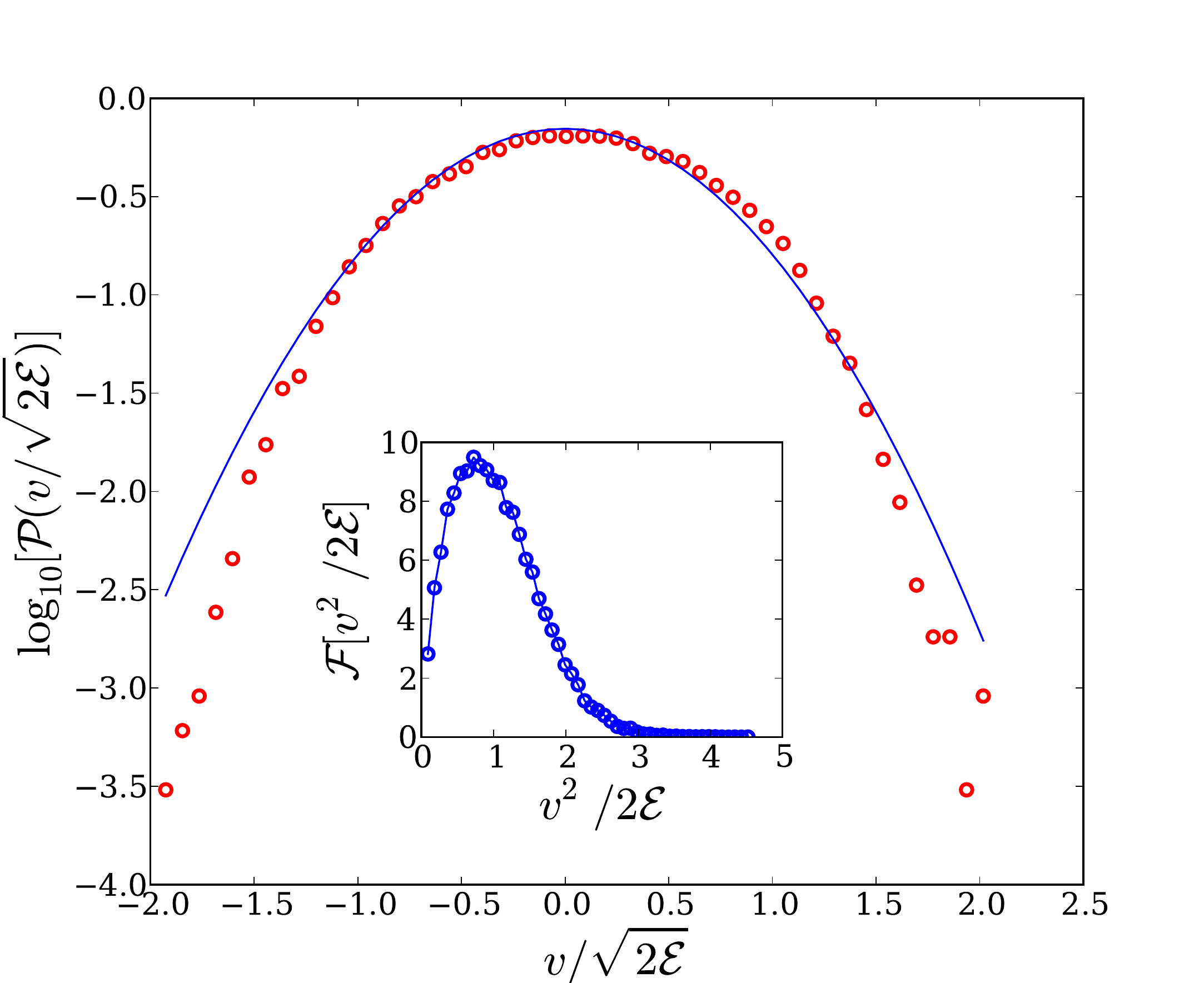}
\caption{\label{fig:lpdf} (Color Online)
Plot of the log of PDF $\mathcal{P}$ of the $x$ component of velocity  of the particles at $\omega t = 100$. The abscissa is normalized by $\sqrt{2\cE}$ at the same time.  Figure~\ref{fig:evst2} shows that by this time $\cE$ has grown by more than five orders of magnitude compared to its initial value. 
The continuous line is the function (parabola), $f(x) = C - x^2/\sigma^2$ with $C=-0.155$ and $\sigma=1.25$.  
The inset shows the PDF $\mathcal{F}$ of kinetic energy ($v^2$)  at the same instant of time. 
}\end{center}
\end{figure}

We further study the PDF of
energies of particles with different random initial conditions;
see inset of Fig.~\ref{fig:lpdf}.  
The PDF of $v_x$ is also plotted in log-lin scale in Fig.~\ref{fig:lpdf}.
For small values of its argument this PDF is well approximated by
a Gaussian, i.e., the PDF of energy would be a Maxwellian,
but for large values of its argument the PDF is sub-Gaussian. 
\begin{table}
\caption{The table shows how the exponent $\xi$ depends on the frequency of the magnetic 
field $\omega$.  We study the limit where the magnetic field changes very slowly. 
For all the runs, $q/m =1$, $\omc=1$ and $\omega/k = 1$. As $\omega/\omc \to 0$,
$\xi$ approached unity.  The values of $\xi$ for smaller $\omega$ has larger error as these runs 
have not ran as long as the first two runs.  }
\label{table:param}
\begin{tabular}{c c c c c}
\hline
\hline
Run            & {\tt R1}    &  {\tt R2}  & {\tt R3}    & {\tt R4}    \\
\hline
$\omega$  & 1/10         & 1/16     &  1/32       & 1/64    \\
$\xi$         &  0.45        &  0.77     &   0.8        &  0.9 \\
\hline
\hline
\end{tabular}
\end{table}

\section{Conclusion}

In this paper we have shown that a helical magnetic field with sinusoidal spatio-temporal dependence
can energize particles.
The energization behavior is a power law in time and, given enough time,
can energize the particles to very high energies where the relativistic effects
start becoming important~\footnote{A relativistic generalization of our model is possible in order to deal 
with ultra-relativistic particle. This is beyond the scope of this paper. }.
In particular, in our simulations we observe 
the mean energy of the particles to grow by six orders of magnitude. 
The chaotic nature of particle trajectories plays a crucial role in our model.
If we change the ABC field such that $A=B=0$ and $C=1$ then we know that
the magnetic lines of force are integrable and non-chaotic.
In such a field, we do observe energization for short times, but at large
times no systematic gain in energy is observed.

The Fermi model of acceleration of charged particle, often referred to as diffusive shock acceleration, is 
thought to be one of the primary mechanisms for energization of charged 
particles in the cosmos, see, e.g., Refs. \cite{dru83,jok01,bal+byk+lin+ray+sch12} for a review.
Fermi's theory also reproduces the experimental observation that the PDF of energies of cosmic rays has 
an inverse power-law tail. 
But diffusive shock acceleration of electrons can occur only if the initial
energy of the electrons is at least of the order of a few MeV which is significantly higher
than the thermal energies; this is the well-known \textit{injection problem}. 

The mechanism we propose is akin to second order Fermi acceleration where a charged particle
is energized due to collisions with random scatter centers moving with random velocities. 
The input ``randomness'' is a crucial ingredient of this process and 
one typically obtains diffusive properties in both real and momentum space \cite{bou+cec+vul04}. 
The Fermi second order process also produces a PDF of energies with power-law tail. 
By contrast, in our case the diffusive behavior is generated by deterministic chaos. 
We obtain sub-diffusive behavior ($\xi < 1$) in momentum space, which becomes close to diffusive
behavior as $\omega\to 0$.  
Furthermore, the  PDF of energies at large times becomes Gaussian with steeper tails, i.e.,
 the PDF can be characterized as Gaussian at low speed, but falls off more rapidly than a Gaussian at
 speeds in excess of the mean.
On the positive side, our model is a possible mechanism that can generate a population of electron
with super thermal energies. They can now act as a resolution to the injection problem. 
What is truly remarkable in our model is that a deceptively simple magnetic field
with a dynamics that is smooth in time is able to energize particles to indefinitely high
energies.

The question of energization of a test (charged) particle in a turbulent plasma has been 
numerically studied in recent times, see, e.g., Refs. \cite{dmi+mat+sen04,ono+isl+vla06}.
These studies consider energization in an electric field that is frozen-in-time but obtained from
one snapshot of a direct numerical simulation of magneto-hydrodynamic turbulence. 
In such a setup, test particles show diffusion in real space. 
Energization is also observed, and at large time energy seems to grow linearly with time. 
In addition, the PDF of energies obtained in Refs.~\cite{dmi+mat+sen04,ono+isl+vla06} has
power-law tails.
However, it is not clear what the exponent of this power-law is and how
that emerges.  Our simple model is able to capture the first two aspects, v.i.z., the random walk and
the energization but not the power-law tail.  One of the contributions to the electric field
in a turbulent plasma comes from the current. The square of the current is the resistive
contribution to energy dissipation rates. 
The energy dissipation rates calculated from the solar wind data~\footnote{
So far, energy dissipation rate has not been directly measured from observations of solar wind turbulence 
but proxies, e.g., average of the cube of the velocity difference across a time scale,  has been used; 
see, e.g., section 7 of Ref.~\cite{mar+tu99} for a review. }
show intermittent behavior. 
Such intermittency is absent in our simple model.  This could be one of the reasons why we do not
observe the power-law tail of the PDF of energy of the test-particles. 
The brings us to the question, what are the minimal
ingredients necessary to add to our model to obtain a power-law tail in the PDF of energy?
This will be the subject of future investigations. 

We thank Caspian Berggren for early contributions to this work,
which was supported in part by the European Research Council
under the AstroDyn Research Project No.\ 227952 (AB and DM)
and the Swedish Research Council under grant 2011-542 (DM).
BD acknowledges supports from NSF grant AGS-1062050 and the 
Individual Investigator Distinguished Research (IIDR) award of the 
University of Alabama in Huntsville.
He is also thankful to Jacob Heerikhuisen and Gang Li
for critical comments and to Gary Zank for his interests and
support. 
EN thanks NORDITA for hospitality. He has been partially supported 
by the Research Academy for Young Scientists (RAYS). 
AR was partially supported by US Department of Energy grant
number DE-FG02-91ER54109.
\printtables


\end{document}